\documentclass{optica-article}
\journal{opticajournal} 

\articletype{Research Article}

\usepackage{lineno}
\usepackage{lipsum}
\begin{document}

\title{Millimeter Wave Imaging using Autler-Townes Splitting Induced Fluorescence from Rydberg Atoms in Rubidium Vapor}

\author{Gour Pati,\authormark{1,*} Daniel Mechael,\authormark{1} Fredrik K. Fatemi,\authormark{2} and Renu Tripathi\authormark{1}}
\address{
\authormark{1}Division of Physics, Engineering, Mathematics \& Computer Science, Delaware State University, Dover, Delaware 19901, USA \\
\authormark{2}DEVCOM Army Research Laboratory, Adelphi, Maryland 20783, USA\\
}
\email{\authormark{*}gspati@desu.edu}

\begin{abstract}
We demonstrate millimeter wave (mmWave) imaging using Autler-Townes (AT) splitting-induced fluorescence in rubidium vapor. By employing counter-propagating probe and coupling beams, we create a plane of atoms in a dark state that is sensitive to incident mmWave and demonstrate the method with mmWaves resonant with the transition between two neighboring Rydberg states at 91.4 GHz. Objects in the mmWave’s path affect the spatial profile of the mmWave field, which, in turn, leads to spatial variation of optical fluorescence at 780 nm. This results in a transduction of the mmWave spatial profile to an optical image that can be acquired by a CCD camera. We demonstrate real-time, diffraction-limited mmWave imaging in transmission geometry, over an effective sensing area of 6.5 cm² with a minimum 600 $\mu{V}$/cm detectable electric field and a rapid 16.4 $\mu{s}$ response time. In addition, we demonstrate this method with a variety of imaging masks, including 3D-printed vortex phase plates. Our study provides a pathway for high-speed and high-resolution mmWave imaging with potential applications in security, communications, and scientific research.
\end{abstract}
\section{\label{sec:level1}INTRODUCTION}
The ability of mmWaves, defined as the range of the electromagnetic spectrum from 30 to 300 GHz, to penetrate materials such as clothing, fog, and smoke, combined with their non-ionizing nature, makes them particularly attractive for defense applications and low-visibility environments \cite{appleby2004,appleby2017}. Millimeter wave (mmWave) imaging has significant potential in a wide range of applications, such as security screening \cite{wang2019, appleby2007,nanzer2012}, nondestructive testing \cite{vakalis2023,murakami2024}, and medical imaging \cite{mukherjee2019,mirbeik2019}. Traditional mmWave imaging methods use Shottky diodes, field effect transistors (FETs), or radiometers coupled to the antenna as a point detector representing a pixel in the mmWave imaging scheme \cite{appleby2017}. Although some of these detectors are fast and sensitive enough for use in active and passive imaging, focal plane array technologies for these detectors have not yet been fully developed \cite{appleby2017,appleby2007}. Also, creating a large focal plane array out of these detectors can become prohibitively expensive and limited in spatial resolution due to the large antenna size. Therefore, mmWave imaging is typically implemented using mechanical (or electrical) scanning of a limited number of detector elements, which requires long data acquisition times \cite{appleby2017}. Computational imaging can help reduce data acquisition time and the number of active components at the expense of an increased computational load, often limiting the resolution and real-time imaging operation \cite{patel2016}.

 Rydberg atoms have emerged as a powerful platform for electrometry. These atoms exhibit high sensitivity to electric field across a wide range of frequencies, ranging from radio waves to terahertz (THz) radiation \cite{fancher2021,zhang2024}. Unlike traditional electro-optical or dipole sensors, Rydberg electrometry offers a quantum-based measurement standard that is inherently traceable to SI units, making them highly reproducible and reliable \cite{fancher2021,zhang2024,holloway2014,tu2024}. Rydberg sensors have been put into applications as receivers for communication applications \cite{meyer2018,deb2018} and mmWave detection \cite{gordon2014,yuan2023,legaie2024,borowka2024}. While most emphasis of Rydberg sensors has been on RF calibration or communications measurements, their use in imaging studies has been limited to a handful of demonstrations \cite{fan2014,wade2017,downes2020,downes2023,schlossberger2024}. Sub-wavelength imaging of variations in 1D microwave electric field strength has been demonstrated by monitoring probe transmission \cite{fan2014}. The conversion of the THz field to the optical field via Rydberg atoms has resulted in a demonstration of THz imaging with high frame rates of up to 3 kHz \cite{wade2017,downes2020,downes2023}. This is achieved by directly monitoring the fluorescence from the Rydberg state. Recent work has shown spatial mapping of external electric fields from few MHz to few GHz and mapping of static magnetic fields by imaging fluorescence from the intermediate 5$P_{3/2}$ state in $^{85}$Rb \cite{schlossberger2024}.

In this paper, we demonstrate active mmWave imaging that leverages Autler-Townes (AT) splitting in a four-level Ladder system involving two Rydberg states to image a resonant mmWave field at 91.4 GHz. Our approach detects fluorescence at 780 nm from the intermediate 5P$_{3/2}$ state in the $^{85}$Rb D2 line to transduce the microwave spatial profile to an optical image. Fluorescence at fixed 780 nm wavelength allows our system to be easily tuned to other Rydberg states for performing mmWave imaging in other bands. Exciting rubidium atoms in a vapor cell using expanded counter-propagating probe and coupling beams in one dimension, we create a plane of atoms in the dark state that exhibits reduced fluorescence due to electromagnetically induced transparency (EIT). When this atomic plane prepared in the dark state is illuminated by a resonant mmWave field, AT splitting reintroduces population into an intermediate state, causing an increase in optical fluorescence. This fluorescence is acquired by a CCD camera. Our approach could offer several advantages over existing mmWave imaging technologies. A response time of 16.4 µs with an active imaging area of approximately 1.3 cm × 5 cm has been achieved, potentially allowing high frame rate operation over 60 kHz. Loss of spatial resolution (or blurring) due to motion of atoms in our approach is calculated using the lifetime of the 5$P_{3/2}$ state to be 6.7 $\mu m$. This is well below the diffraction-limited resolution of microwaves (i.e. 3.28 mm for 91.4 GHz in our demonstration). We demonstrate a sensitivity of approximately 0.6 $\frac{mV}{cm\sqrt{Hz}}$, which makes our approach suitable for active, stand-off mmWave imaging.

This paper is organized as follows. In Sec. II, we provide theoretical description and numerical simulations of the four-level Rydberg system to explain AT-splitting-induced fluorescence, which is used as the basis for mmWave imaging. Sec. III outlines our experimental setup. In Section IV, we present experimental results and characterizations of spatial resolution, response time, and sensitivity of our approach. In this section, we also discuss the fundamental sensitivity limits of this approach along with its current limitations and proposed future improvements to enhance imaging performance. In Sec. V, we conclude with a summary of our findings and their broader implications on mmWave imaging.

\section{\label{sec:level2}THEORETICAL ANALYSIS}
\begin{figure}
\centering
\includegraphics[width=0.5\textwidth]{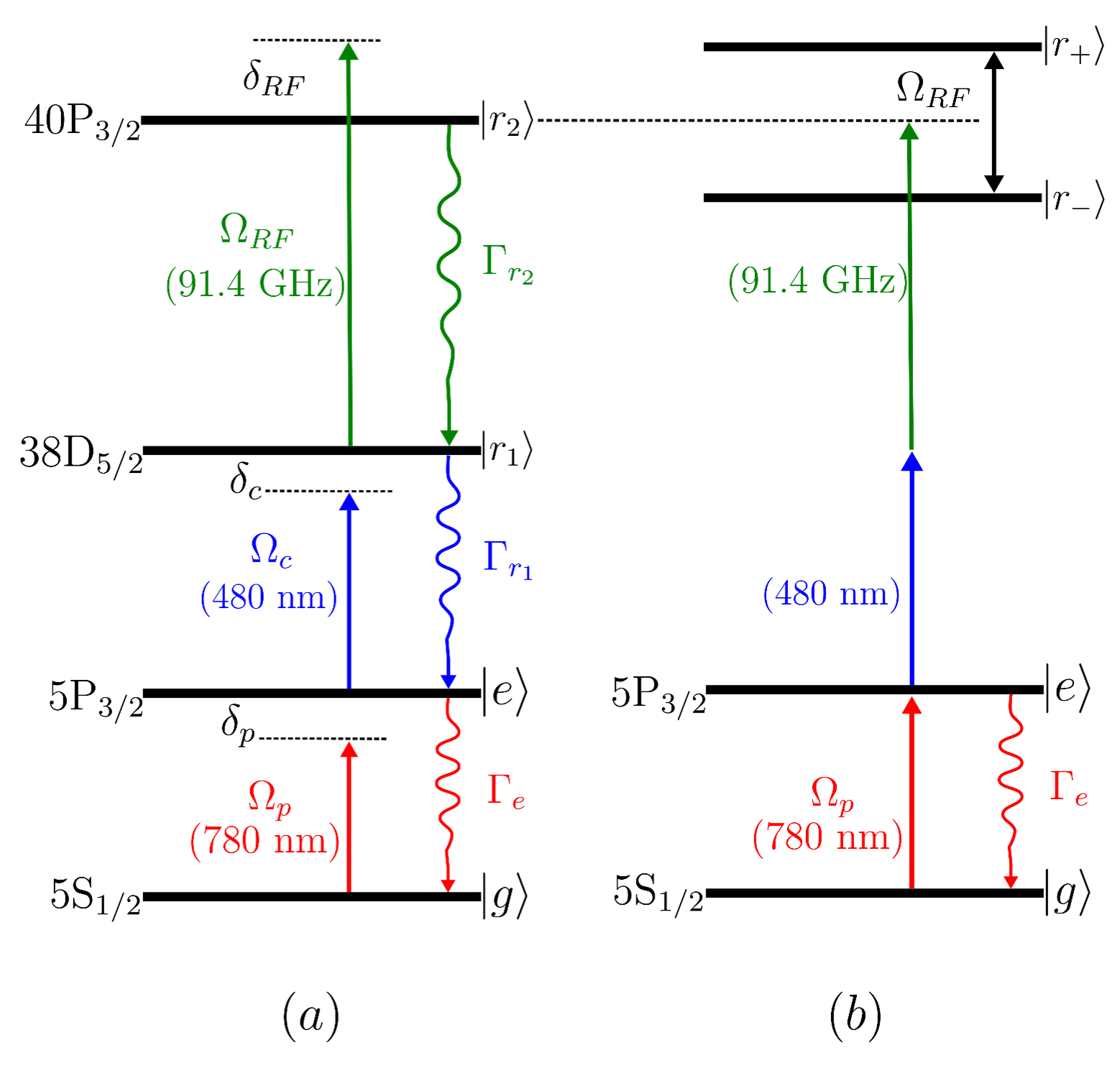}
\caption{\label{fig:epsart} (a) Energy level diagram of four-level Ladder system in $^{85}Rb$ D1 manifold with relevant couplings, decay rates, and detunings. (b) Energy level diagram of the resonant four-level system with Rydberg states dressed by the RF field. Splitting between the new light shifted Rydberg states is $\Omega_{RF}$.}
\label{fig1}
\end{figure}

In this section, we describe the framework of a four-level Ladder system in the $^{85}$Rb D1 manifold used in mmWave imaging. We outline the population dynamics first, in the resonant three-level EIT system and then describe the AT splitting regime in a four-level system, showing how the presence of the resonant mmWave field alters the steady-state population distribution. This forms the basis of our imaging technique, which relies on detecting changes in 780 nm fluorescence as an indicator of the mmWave field strength. The system under investigation consists of four key atomic states in $^{85}$Rb in a series of optical and mmWave (or RF) transitions between them, as shown in Fig. \ref{fig1}(a). The ground state, $|g\rangle$, corresponds to the F = 3 level in the 5S$_{1/2}$ state. It is coupled to the intermediate excited state, $|e\rangle$ (i.e. F = 4 level in the 5$P_{3/2}$ state) by a probe beam at wavelength, $\lambda_p\simeq$ 780 nm. This coupling is characterized by the probe Rabi frequency $\Omega_p$. 

The first Rydberg state in the system, $|r_1\rangle$ is the $38D_{5/2}$ state, which is coupled to the intermediate state by a coupling beam at approximately wavelength, $\lambda_c \simeq$ 480 nm with a corresponding Rabi frequency $\Omega_c$. Lastly, the second Rydberg state, $|r_2\rangle$ corresponds to the 40$P_{3/2}$  state, which is coupled to $|r_1\rangle$ via the incident mmWave imaging field at 91.4 GHz, characterized by the RF Rabi frequency $\Omega_{RF}$. A strong RF coupling occurs at a relatively small mmWave field due to the high transition dipole moment between the two Rydberg states. The presence of this strong RF coupling between $|r_1\rangle$ and $|r_2\rangle$ Rydberg states leads to new light-shifted eigenstates $|r_\pm \rangle$, which are superpositions of $|r_1\rangle$ and $|r_2\rangle$, as shown in Fig. \ref{fig1}(b). These states are defined by $|r_{\pm} \rangle = \frac{1}{\sqrt{2}}  [|r_1\rangle \pm |r_2\rangle]$. These light-shifted eigenstates, referred to as dressed states, have an energy difference equal to $\hbar\Omega_{RF}$, which is directly related to the strength of the RF electric field. Consequently, as the probe beam detuning, $\delta_P$ (or coupling beam detuning, $\delta_C$) is scanned, the system comes into two-photon resonance with these RF-dressed eigenstates, leading to two AT split EIT transmission peaks, separated by $\Omega_{RF}$. Because this splitting is proportional to the RF electric field strength, the AT splitting has become a common method of RF electrometry through the THz regime, offering excellent sensitivity and accuracy while being easily traceable back to SI units [16].

We model the four-level Ladder system described above using the Lindblad master equation \cite{manzano2020} given by
\begin{equation}
\dot{\rho} = \frac{i}{\hbar} [\rho, H] + \sum_{j=2}^{4} \Gamma_j \left( L_j \rho L_j^{\dagger} - \frac{1}{2} \{ L_j^{\dagger} L_j, \rho \} \right)
\end{equation}
where $\rho$ represents the density matrix. $H$ is the total Hamiltonian in the rotating frame, described as
\begin{equation}
H = \frac{\hbar}{2}
\begin{bmatrix}
0 & \Omega_p & 0 & 0 \\
\Omega_p & -2\delta_p & \Omega_c & 0 \\
0 & \Omega_c & -2(\delta_p + \delta_c) & \Omega_{RF} \\
0 & 0 & \Omega_{RF} & -2(\delta_p + \delta_c + \delta_{RF})
\end{bmatrix}
\end{equation}

The decay operators $L_j$ represent all the decay paths in the system with $\Gamma_j$  ($\Gamma_2=\Gamma_e,\Gamma_3=\Gamma_{r_1}$, and $\Gamma_4=\Gamma_{r_2 }$) being the decay rate of the indicated path
\begin{equation}
L_j  = |j-1\rangle\langle{j}|
\end{equation}

\begin{figure}
\centering
\includegraphics[width=0.6\textwidth]{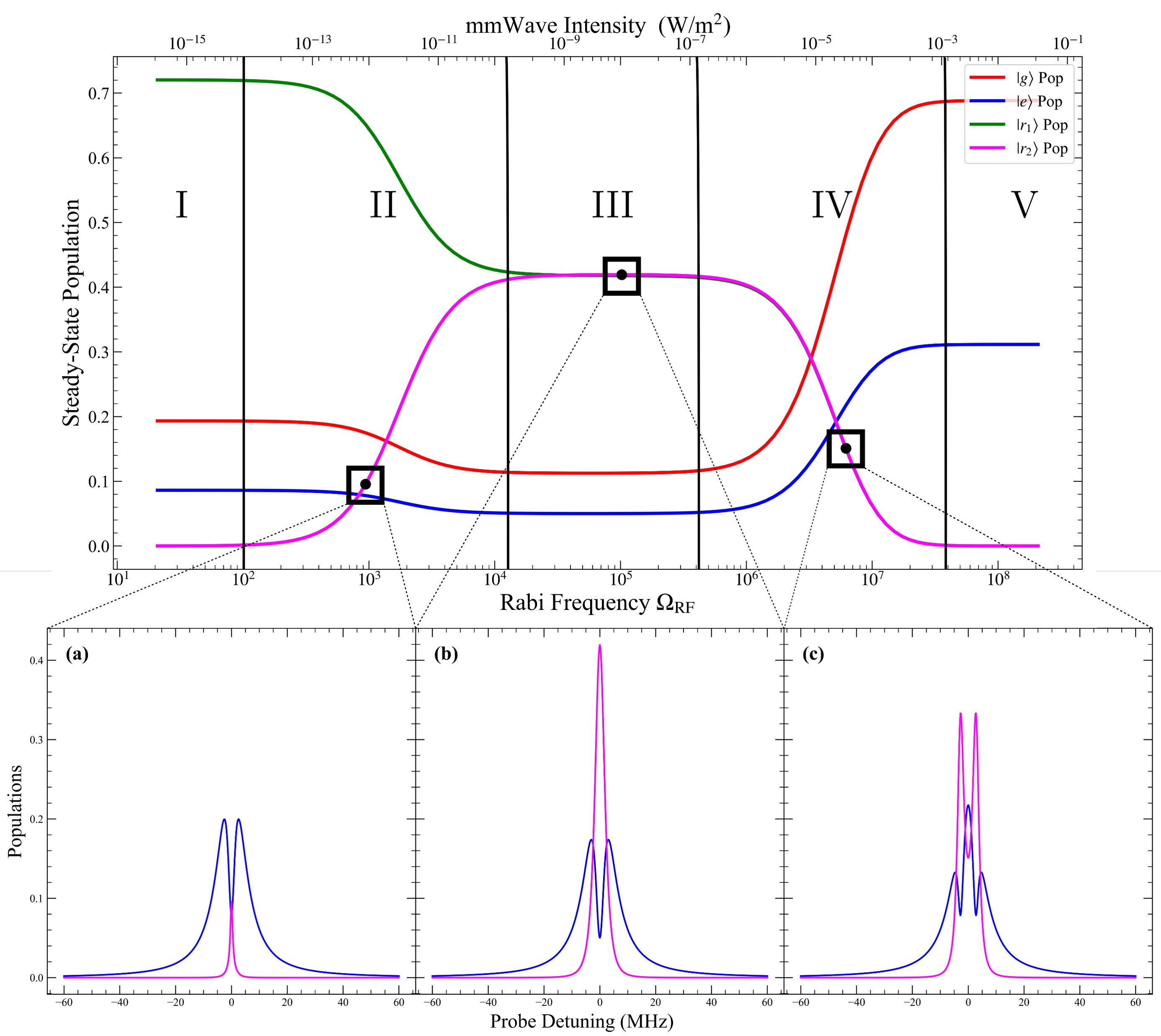}
\caption{\label{fig:epsart} The top plot shows steady-state populations of the four-level Ladder system as a function of the RF Rabi frequency, $\Omega_{RF}$ (in log scale), highlighting five distinct dynamical regions (I-V). A probe detuning scan is shown in plots (a), (b), and (c) below, for the three points indicated within regions II, III, and IV respectively. Plot (a) shows resonances in the population of states $|e\rangle$ and $|r_2\rangle$ occuring in region II with dynamics corresponding to a pseudo two-level system formed between the Rydberg states $|r1\rangle$ and $|r2\rangle$. Plot (b) shows resonances occurring in region III which is designated as the saturation plateau. Plot (c) shows resonances occurring in region IV which is designated as the AT splitting regime where the light-shifted EIT resonance leads to single-photon absorption and, thus, increased fluorescence at zero detuning.}
\label{fig2}
\end{figure}

To analyze the system’s resonant response to varying mmWave intensity, we performed a numerical simulation using the RydIQule Python package \cite{miller2024}. The Rabi frequencies, $\Omega_p$ and $\Omega_c$, were calculated based on the experimental parameters for laser power and the beam area. Atomic parameters such as decay rates, transition dipole moments, and state energies were calculated using the ARC: Alkali Rydberg Calculator \cite{sibalic2017}. For our simulations, we used the following parameters: $\Omega_p$  = 4 MHz, $\Omega_c$  = 0.3 MHz, $\Gamma_2/2\pi$ = 6 MHz, $\Gamma_3/2\pi$ = 2.9 kHz, and $\Gamma_4/2\pi$ = 1.2 kHz. The top plot in Figure 2 shows the populations of the four atomic states as a function of $\Omega_{RF}$ obtained under the steady-state (SS) condition, i.e. $\dot{\rho}$ = 0. This plot is divided into five distinct regions, described below according to the relevant dynamics:

\textbf{Region I (Weak RF Field):} This region in Fig. \ref{fig2} is referred to as the weak RF field region for which $\Omega_{RF}$  $\ll$ $\Gamma_{r_2}$. In this case, the Rydberg state $|r_2\rangle$ is essentially uncoupled from the three-level system (formed by the states $|g\rangle$, $|e\rangle$, and $|r_1\rangle$, as shown in Fig. 1), for which there is a well-defined SS solution, referred to as the dark state $|-\rangle$, which is a coherent superposition of the $|g\rangle$ and $|r1\rangle$ states, i.e.

\begin{equation}
|-\rangle = \frac{\Omega_c |g\rangle - \Omega_p |r_1\rangle}{\sqrt{\Omega_p^2 + \Omega_c^2}}
\end{equation}

This dark $|-\rangle$ state is completely decoupled from the intermediate state, $\langle{e}|H|-\rangle$ = 0, leading to minimal population in the intermediate state, $|e\rangle$. In an ideal three-level system, the population of $|e\rangle$ would be reduced to zero. However, the de-phasing of the dark state, caused by the decay rate, $\Gamma_{r_1}$, allows some population to remain in $|e\rangle$. Our system remains in this condition while the mmWave field is off, producing minimal fluorescence from $|e\rangle$. This is considered as background fluorescence.

\textbf{Region II (Moderate RF Field):} In this region of Fig. \ref{fig2}, $\Omega_{RF}$  is comparable to $\Gamma_{r_2}$. Here, our simulation shows that the population is being transferred from all three states $|g\rangle, |e\rangle,$ and $|r_1\rangle$ into $|r_2\rangle$, most noticeably from $|r_1\rangle$. Treating $|r_1\rangle$ and $|r_2\rangle$ as a pseudo two-level system, we find this region to eventually saturate with further increase in $\Omega_{RF}$. Plot \ref{fig2}(a) shows the resonances in population of states $|e\rangle$ and $|r_2\rangle$ observed as a function of probe detuning $\delta_p$ for a fixed value of $\Omega_{RF}$ shown as a point within the square in region II. While the population of state $|e\rangle$ does not look appreciably different from the three-level system under the SS condition, we see that population is transferred into $|r_2\rangle$ on three-photon resonance. Region II was chosen as a Rydberg-based imaging approach in which THz imaging is performed by measuring the fluorescence directly from the Rydberg state. \cite{wade2017,downes2020,downes2023}.

\textbf{Region III (Saturation Plateau):} Following the simplistic description of the pseudo two-level system, our simulation ultimately shows a saturation plateau where the populations of $|r_1\rangle$ and $|r_2\rangle$ equalize. Further increase in $\Omega_{RF}$ does not lead to any changes in the SS population. The resonance plots in Fig. \ref{fig2}(b) show that the resonance in the population of state $|e\rangle$ has not changed appreciably over this region compared to region II. However, $|r_2\rangle$ has gained considerable population on three-photon resonance.

\textbf{Region IV (Autler-Townes Splitting):} Further increasing $\Omega_{RF}$ in a true two-level system would not lead to any change in the SS population. However, the RF coupling dresses the two Rydberg states $|r_1\rangle$ and $|r_2\rangle$, creating two new eigenstates with energy separation of $\hbar\Omega_{RF}$, as shown in Fig. \ref{fig1}(b). Splitting of these eigenstates is referred to as AT splitting. These dressed states are no longer two-photon resonant with the transition involving the $|g\rangle$ and $|e\rangle$. Because the system is no longer two-photon resonant at zero detuning \cite{anisimov2011}, single-photon absorption is reintroduced between $|g\rangle$ $\rightarrow$ $|e\rangle$ leading to increased population in $|e\rangle$, and thereby, increased fluorescence from, $|e\rangle$. AT splitting becomes visible when it becomes comparable to the EIT linewidth, $\Gamma_{EIT}$. When we scan the frequency of coupling (or the probe) beam around zero detuning, two transmission peaks will appear as the system becomes resonant with each of the dressed Rydberg eigenstates. Figure \ref{fig2}(c) demonstrates this phenomenon by showing increased population in $|e\rangle$ on three-photon resonance and shifted resonance peaks corresponding to the new eigenstates with peaks located at $\delta_p$  = $\pm \frac{\Omega_{RF}}{2}$. As the peaks continue to separate, single-photon absorption from $|g\rangle$ $\rightarrow$ $|e\rangle$ reemerges and continues to increase.

\textbf{Region V (Saturated AT Splitting):} Once the AT-split resonances no longer overlap at zero detuning, the population in $|e\rangle$ has reached its maximum and is saturated. This region could remain useful for electrometry by measuring the AT splitting. However, as an increase in $\Omega_{RF}$ does not increase the population in $|e\rangle$ (or fluorescence resulting from $|e\rangle$), it makes this region unsuitable for mmWave imaging.

Our imaging scheme operates in Region IV where increasing $\Omega_{RF}$ leads to a proportional increase in the population of $|e\rangle$ and thereby, increasing the fluorescence resulting from the decay of atoms from $|e\rangle$ to $|g\rangle$. We produce a mmWave sensitive screen that is resonant with the $|r_1\rangle$ $\rightarrow$ $|r_2\rangle$ transition by exciting atoms with expanded and counter-propagating probe and coupling beams, and creating a plane of atoms (in a vapor medium) prepped in the dark state described in Eqn. (4). By imaging the incident mmWave onto this screen, and then subsequently imaging the optical fluorescence from $|e\rangle$ $\rightarrow$ $|g\rangle$  using a near infrared (NIR) sensitive CCD camera, we image the mmWave field. It is also important to note that the simulations shown in Fig. \ref{fig2} were performed without accounting for the motion of the atoms. We extended our simulation by including velocity averaging in Eqn. (1) and found that in the velocity-averaged SS solution, the five regions described in the stationary case remain visible although the atomic populations in the states $|e\rangle$, $|r_1\rangle$ and $|r_2\rangle$ are significantly diminished. This is due to the fact that the Doppler shifted detunings $\delta_p$ and $\delta_c$ for most velocity groups become non-zero to create two-photon excitation, which leads to majority of the population in the ground-state $|g\rangle$. We also found that the location of Region IV that is relevant to our mmWave imaging scheme shifts to a higher  $\Omega_{RF}$ value due to velocity-induced broadening of the EIT linewidth. Another point to note is that mmWave imaging using our scheme can also be performed with detuned (or off-resonant) RF excitation. In this case, fluorescence produced from Rydberg atoms will be influenced by AC Stark shift rather than AT splitting as in the resonant case \cite{yao2022}.

\section{\label{sec:level2}EXPERIMENTAL DESCRIPTION}
Figure 3 shows a three-dimensional layout of our experimental mmWave imaging setup. The probe beam (red color) at 780 nm is generated by a tunable diode laser (Toptica DL Pro) that is stabilized to the 5$S_{1/2}$,F = 3 $\rightarrow$ 5$P_{3/2}$,F = 4 transition using saturated absorption spectroscopy in a rubidium reference cell. The coupling beam (blue color) at 480 nm is produced by a high-power, tunable, frequency-doubled laser (Toptica TA-SHG Pro). This beam is scanned across the 5$P_{3/2}$  $\rightarrow$ 38$D_{5/2}$ Rydberg transition and locked to the EIT peak observed in a second rubidium reference cell. The counterpropagating probe and coupling beams are expanded and collimated into a uniform line and aligned through the experimental rubidium cell of length 5 cm and diameter 2.5 cm. In our case, beam shaping for both beams is accomplished using $5^o$ angle Powell lenses which expand the beams along the vertical dimension, followed by  cylindrical lenses with 150 mm focal length to create a collimated beam with a vertical dimension of $\approx$ 1.3 cm and horizontal dimension of $\approx$ 350 $\mu$m FWHM. Before expansion, the optical power used in the probe beam is $\approx$ 1.6 mW and $\approx$ 500 mW in the coupling beam. As we discussed in Sec. II, the two counter-propagating, expanded, and collimated laser beams create a sheet of atoms in the rubidium cell participating in the EIT process.
\begin{figure}
\centering
\includegraphics[width=0.7\textwidth]{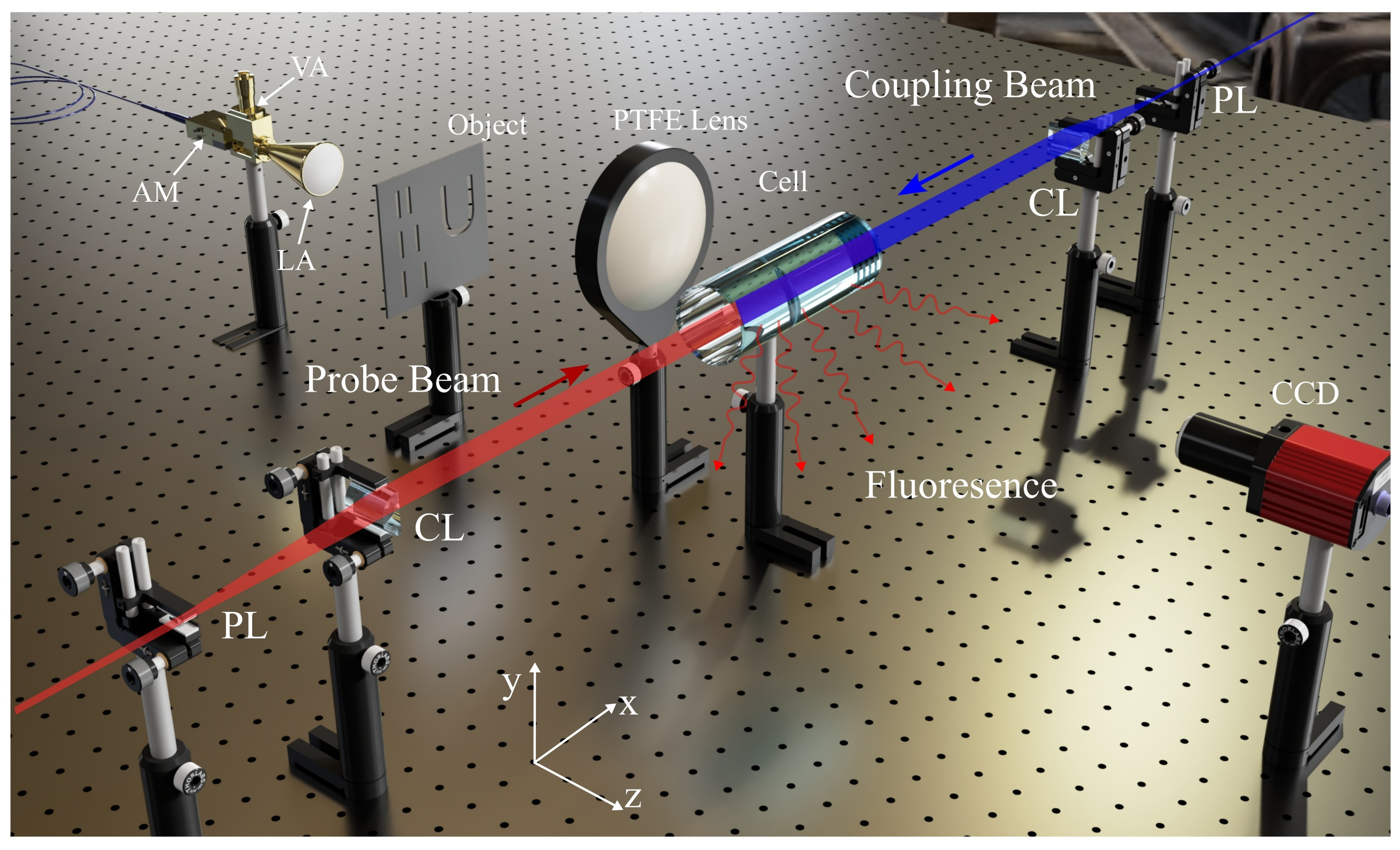}
\caption{\label{fig:epsart} A three-dimensional layout of the mmWave imaging setup described in this work. Counter-propagating probe and coupling beams are shaped into a light sheet to create two-photon excitation inside a rubidium vapor cell. Millimeter wave image of the object is converted to an optical image using AT splitting induced fluorescence produced from Rydberg atoms. The choice of coordinate axes is indicated in the diagram. PL: Powell lens; CL: cylindrical lens AM: active multiplier (8x); VA: variable attenuator; LA: lens antenna.}
\end{figure}

The mmWave field at 91.4 GHz, resonant with the 38$D_{5/2}$  → 40$P_{3/2}$ transition, is generated using an active 8x frequency multiplier. The input to this multiplier is provided by a microwave signal generator (Anritsu, MG3692C), set to 11.425 GHz. The output mmWave field is collimated into a Gaussian beam with a waist of approximately 1.9 cm using a collimating lens antenna (LA) shown in Fig. 3. A 30 dB variable attenuator is used to control the mmWave power. Excitation of the atom sheet in the rubidium cell with the mmWave creates AT-splitting-induced fluorescence, as discussed in Sec. II. For imaging, the mmWave antenna is positioned perpendicular to the atom sheet created in the rubidium cell. The object to be imaged, is placed in the path of the mmWave beam as shown in Fig. 3. The mmWave field transmitting through and diffracted (or scattered) by this object, is imaged onto the atom sheet with a demagnification factor corresponding to a polytetrafluoroethylene (PTFE) lens (50mm diameter, 40 mm focal length) between the object and the rubidium cell, as shown in Fig. 3. Regions with higher mmWave field strength induce AT splitting in the atom sheet, thus increasing the fluorescence at 780 nm. This effect causes a transduction of the mmWave image to an optical image which is acquired by a CCD camera. Depending on the required resolution and exposure time, we used two separate cameras for optical imaging. For high-speed imaging with short exposure times below 20 ms, a 1.2 MP CMOS camera (Thorlabs, CS135MUN) is employed due to its high NIR sensitivity. For higher-resolution imaging at longer exposure times (up to 200 ms), a 20 MP CMOS camera (Tucsen, FL 20BW) is used. In both cases, a bandpass filter (with 3 nm bandwidth centered at 780 nm) is placed in front of the camera to suppress any stray light. Background subtraction is also used to improve contrast. This is done in real-time by first acquiring the background image by turning the mmWave field off and then doing the subtraction during mmWave image acquisition. We first characterized the spatial resolution, temporal response, and sensitivity of our system before demonstrating mmWave imaging. These are discussed in the following section.

\section{\label{sec:level2}RESULTS \& DISCUSSIONS}

Initially, we characterized the mmWave beam profile to get information about the spatial resolution imposed by the mmWave imaging optics. The collimated mmWave is focused through a PTFE lens, and the fluorescence from the rubidium cell placed at the focus is imaged onto the CCD camera. The image, shown in Fig. 4(a), exhibits an Airy pattern, with a bright central spot surrounded by diffraction rings.
\begin{figure}
\centering
\includegraphics[width=0.7\textwidth]{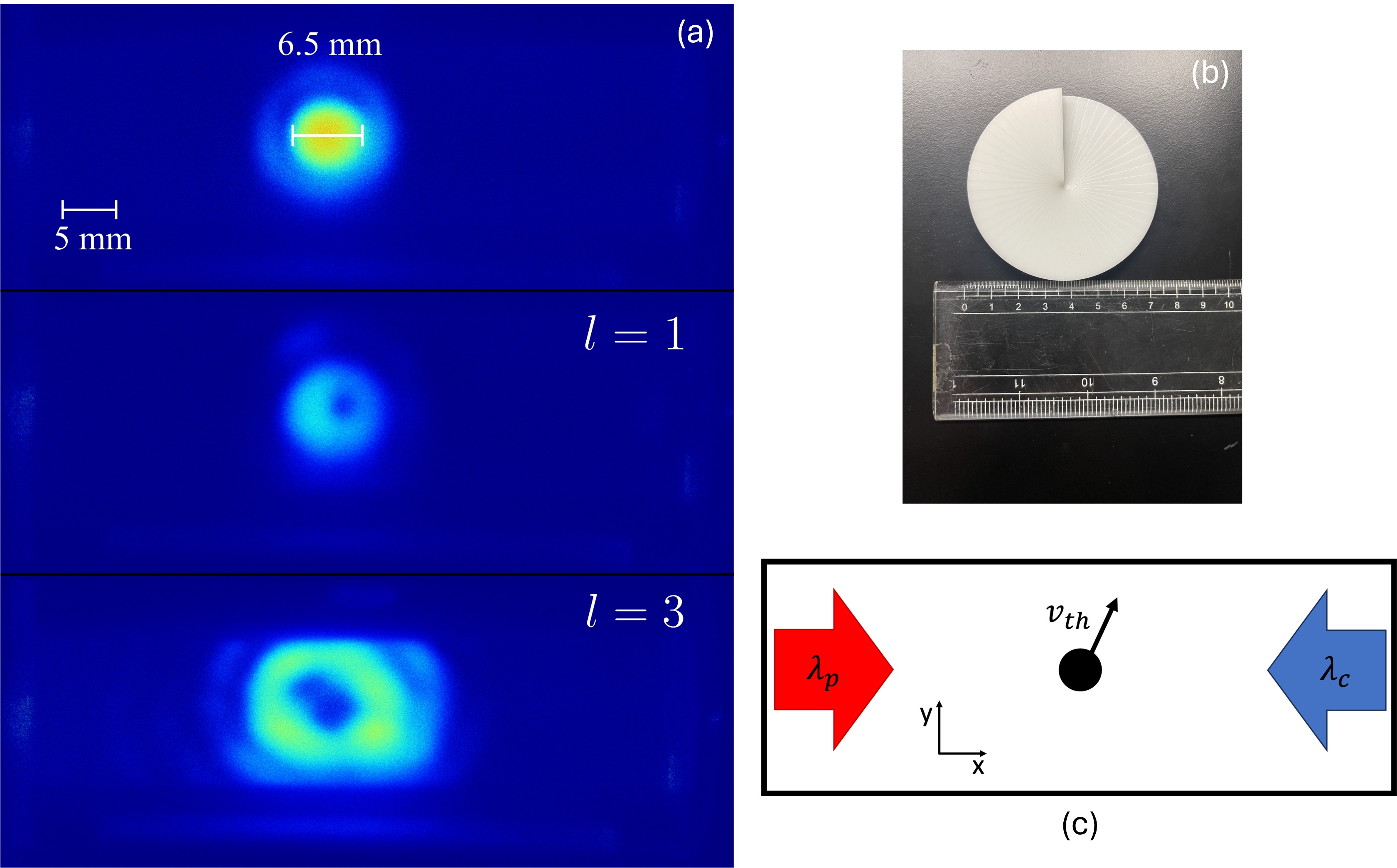}
\caption{\label{fig:epsart} Millimeter wave imaging in the focal plane using AT-splitting-induced fluorescence from Rydberg atoms in a rubidium cell. (a) TOP: Airy diffraction pattern produced by PTFE lens at the focal plane. MIDDLE: OAM imaging using $l$ = 1 SPP shown in (b). Vortex is produced due to phase singularity introduced by SPP. BOTTOM: OAM imaging using $l$ = 3 SPP. Images are acquired using the Tucsen CCD camera with 200 ms exposure time, and image size is calibrated using the camera. (c) Diagram illustrating the atomic motion within the light sheet defined in the (xy) plane. Arrows indicate propagation directions of wavelength mismatched probe and coupling beams, and thermal motion of the atom, in general, with respect to these beams.}
\end{figure}
The size of this central focal spot (defined here as the diameter of the first Airy pattern minimum) is measured to be approximately 6.4 mm, which agrees well with the diffraction-limited spatial resolution (i.e. 6.4 mm) of our system given by the relation: d = 2.44$\lambda_{RF}\frac{f}{D}$, where $\lambda_{RF}$ is the mmWave wavelength ($\simeq$ 3.28 mm), and $\frac{f}{D}$ corresponds to the f-number (i.e. f$\#$ = 0.8) of the PTFE lens. We used our imaging geometry to image mmWave carrying OAM information. For this purpose, we designed a few spiral phase plates (SPPs) to impart OAM to the mmWave. These phase plates are 3D printed using a commonly available UV-curable SLA resin\cite{vargas2022,sahin2019,schemmel2014}. Figure 4(b) shows picture of a 3D printed SPP which is designed to produce the lowest-order OAM beam with integer value $l$=1. For OAM imaging, the phase plate is introduced into the mmWave beam path by positioning it in the front focal plane of the PTFE lens. The resulting intensity distributions for $l$=1 and $l$=3 SPPs in Fig. 4(a: middle and bottom) show vortex structures with dark central regions, corresponding to phase singularities induced by these phase plates \cite{downes2022}. The size of the dark center increases with increasing $l$ value (or topological charge). We observed distortions in the shape of the dark center for $l$=3 SPP in Fig. 4(a: bottom). This is due to imperfections in designing the spiral phase structure for $l$=3. Millimeter wave orbital angular momentum (OAM) beams could find potential applications in mobile communications as they allow higher data transfer rates by multiplexing different OAM modes \cite{ge2017,yan2014}. Our results here demonstrate a simple mechanism for real-time mmWave OAM beam readout which could be relevant for applications in communication.

In the preceding paragraph, we discussed the spatial resolution limit imposed by the PTFE lens and the microwave wavelength. There is another relevant limit to spatial resolution due to the motion of atoms. As discussed in Sec. II, in the presence of a mmWave field, the AT splitting allows for single-photon absorption from $|g\rangle$ to $|e\rangle$ at 780 nm and emission of these photons in fluorescence through decay from the intermediate state $|e\rangle$. During this absorption-emission cycle, the atom moves an average distance of $v_{th} \tau_{|e\rangle}$ = 6.5 $\mu{m}$, where $v_{th}$ = $\sqrt\frac{k_B T}{m}$ $\approx$ 170 m/s is the average thermal velocity of atom at room temperature (i.e. $T$ $\simeq$ 300 K), $m$ is the mass of Rb atom, $k_B$ is the Boltzmann constant, and $\tau_{|e\rangle}$ is the lifetime of $|e\rangle$ ($\approx$ 26 ns). This atomic motion creates a loss of spatial resolution for the atom sheet in the y-direction [i.e. the direction perpendicular to the propagation of light beams, as shown in Fig. 3(c) and Fig. 4(c)] in which atoms do not experience the Doppler effect. In the x-direction [i.e. along the direction of propagation of the light beams], the wavelength mismatch between the probe and coupling beams leads to highly unequal Doppler shifts, restricting the velocity groups that participate in EIT to $v_{max}$ = $\Gamma_{EIT}$ $\frac{\lambda_p \lambda_c}{(\lambda_p-\lambda_c)} \simeq$ 12.48 m/s. This velocity is calculated using the experimentally measured EIT linewidth, $\Gamma_{EIT}$ $\simeq$ 10 MHz. We determined the motion-induced blurring in the x-direction to be $v_{max}$$\tau_{|e\rangle}$ $\approx$ 0.3 $\mu{m}$. Thus, motion-induced blurring in both x- and y-directions are negligible compared to the diffraction-limited imaging resolution imposed by the microwave wavelength and the PTFE lens.

\begin{figure}
\centering
\includegraphics[width=0.7\textwidth]{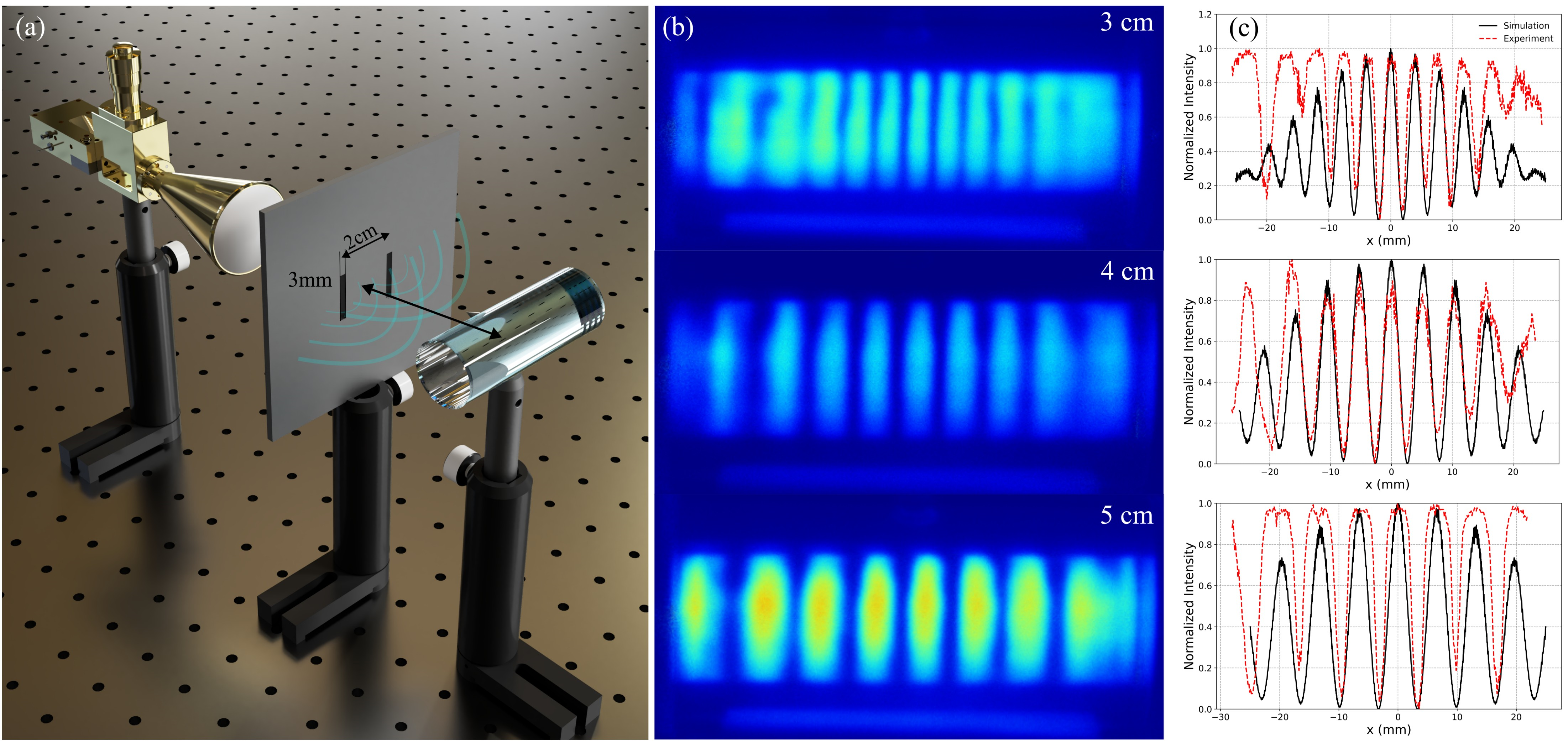}
\caption{\label{fig:epsart} (a) Millimeter wave imaging configuration used to image double-slit diffraction pattern. (b) Pictures of diffraction patterns acquired using a CCD camera, with the double-slit positioned 3, 4, and 5 cm away from the cell, respectively. (c) Line traces (red color) showing intensity distributions across the diffraction patterns shown in (b) and corresponding comparisons with simulations (black color). Interference fringes are observed, and the spatial frequencies of these interference fringes match closely with simulations.}
\end{figure}

Next, we demonstrate the system’s ability to image the double-slit diffraction pattern without a lens. We chose a double-slit made using a 4 mm thick aluminum plate with a slit width of 3 mm and a center to center separation of 2 cm. Figure 5(a) shows the imaging geometry. The diffraction pattern produced by the double-slit illuminates the rubidium cell and is imaged onto the CCD camera (not shown) using fluorescence emitted from the cell. Figure 5(b) shows pictures of near-field diffraction patterns produced by the double-slit positioned at 3, 4, and 5 cm away from the cell, respectively. These pictures of near-field diffraction patterns are acquired over the entire length of the cell (i.e. 5 cm). Interference fringes are observed in the diffraction patterns. Figure 5(b) shows reduction in the fringe-density with increasing distance between the cell and the double-slit. We used a beam propagation simulation (BeamLab: Matlab Toolboxes \cite{BeamLab}) to verify these diffraction patterns produced by the mmWave. Figure 5(c) shows intensity distributions in red color (normalized using the maximum gray value) obtained by taking lengthwise line traces closer to the center of diffraction patterns shown in Fig. 5(b) and corresponding matches with simulation shown in solid dark lines. Figure 5(c) shows good agreement between simulation and experiment in all three cases, particularly in terms of the number of fringes observed across the field of view. However, we observed deviations in experimental traces from simulation towards the edges, which are caused likely due to reflections of mmWave from the cell walls. 
\begin{figure}
\centering
\includegraphics[width=0.7\textwidth]{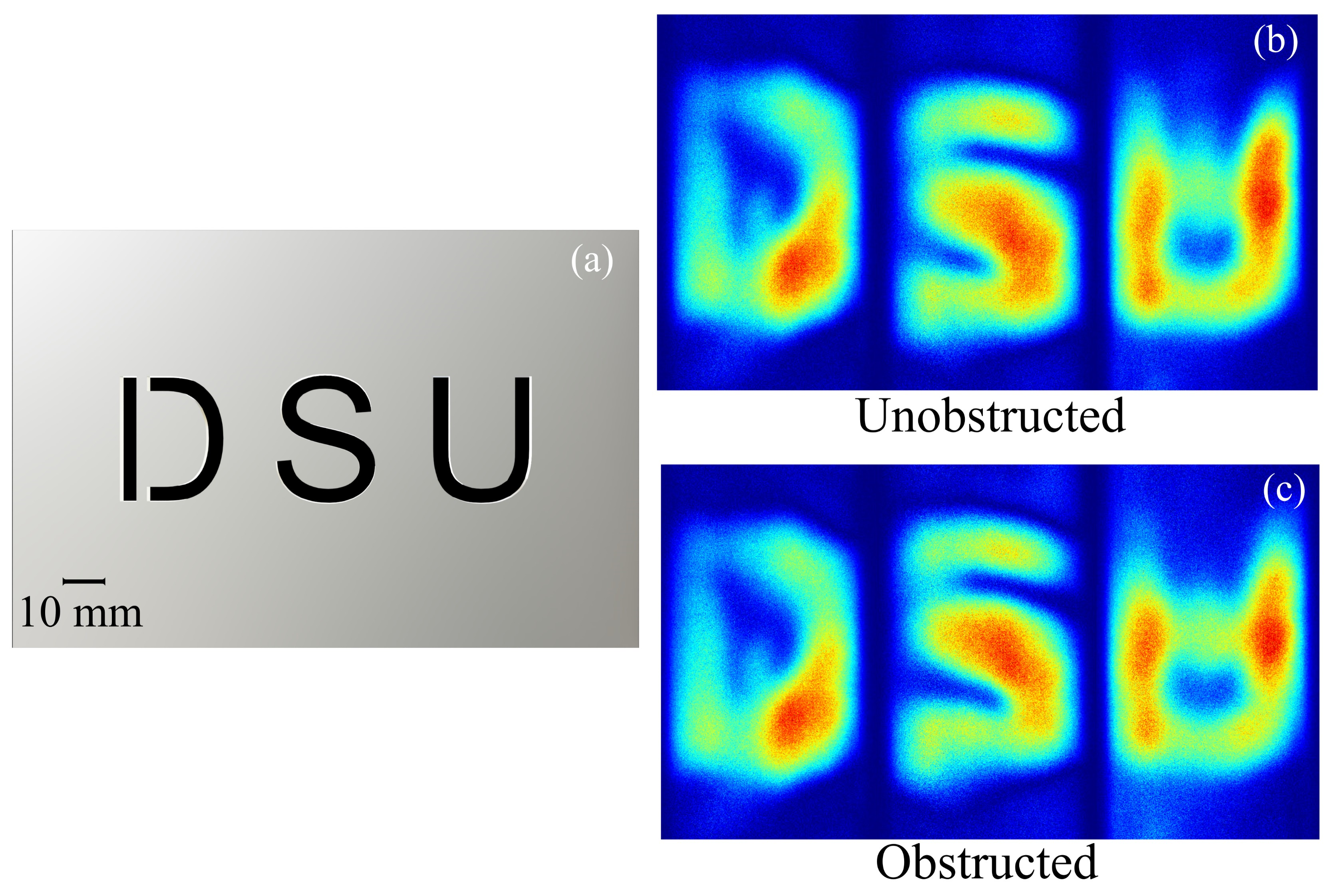}
\caption{\label{fig:epsart} (a) DSU metal stencil with size indicated by the scale bar, (b) Millimeter wave image acquired using fluorescence from Rydberg atoms onto the CCD camera, and (c) obstructed image acquired by placing a 5 mm thick cardboard in front of the metal stencil.}
\end{figure}

To demonstrate direct mmWave imaging, we imaged a metal ‘DSU’ stencil, shown in Fig. 6(a), using the PTFE lens in a transmission geometry. In this case, the metal object is illuminated from behind by the collimated mmWave (produced by the lens antenna as shown in Fig. 3) and the transmitted/diffracted wave is imaged onto the rubidium cell by the lens with appropriate demagnification. Fluorescence produced by Rydberg atoms converts the mmWave image to an optical image which is acquired by a CCD camera. Images of ‘DSU’ stencil are shown in Figs. 6(b,c). Because of the limited size of the antenna beam, we could not illuminate the entire object simultaneously to produce this image. Instead, the letters in the stencil were sequentially illuminated and the individual images stitched together to form the full image of the stencil. The images were captured using the Tucsen CCD camera and a 200 ms exposure. Background subtraction was used in post-processing to improve image contrast. Although we are able to produce interpretable images, the image quality is affected by the coherent mmWave illumination which gives rise to speckle artifacts. The features on the stencil are small and have sizes comparable to the wavelength of mmWave, causing significant diffraction and diffusivity across the edges leading to blurring. The image shown in Fig. 6(c) is acquired with an obstruction created by placing a 5mm thick piece of cardboard between the stencil and the PFTE lens. Despite this obstruction, no significant change in the image is observed. This illustrates the penetration capability of mmWaves in this frequency range.
\begin{figure}[h]
\centering
\includegraphics[width=0.6\textwidth]{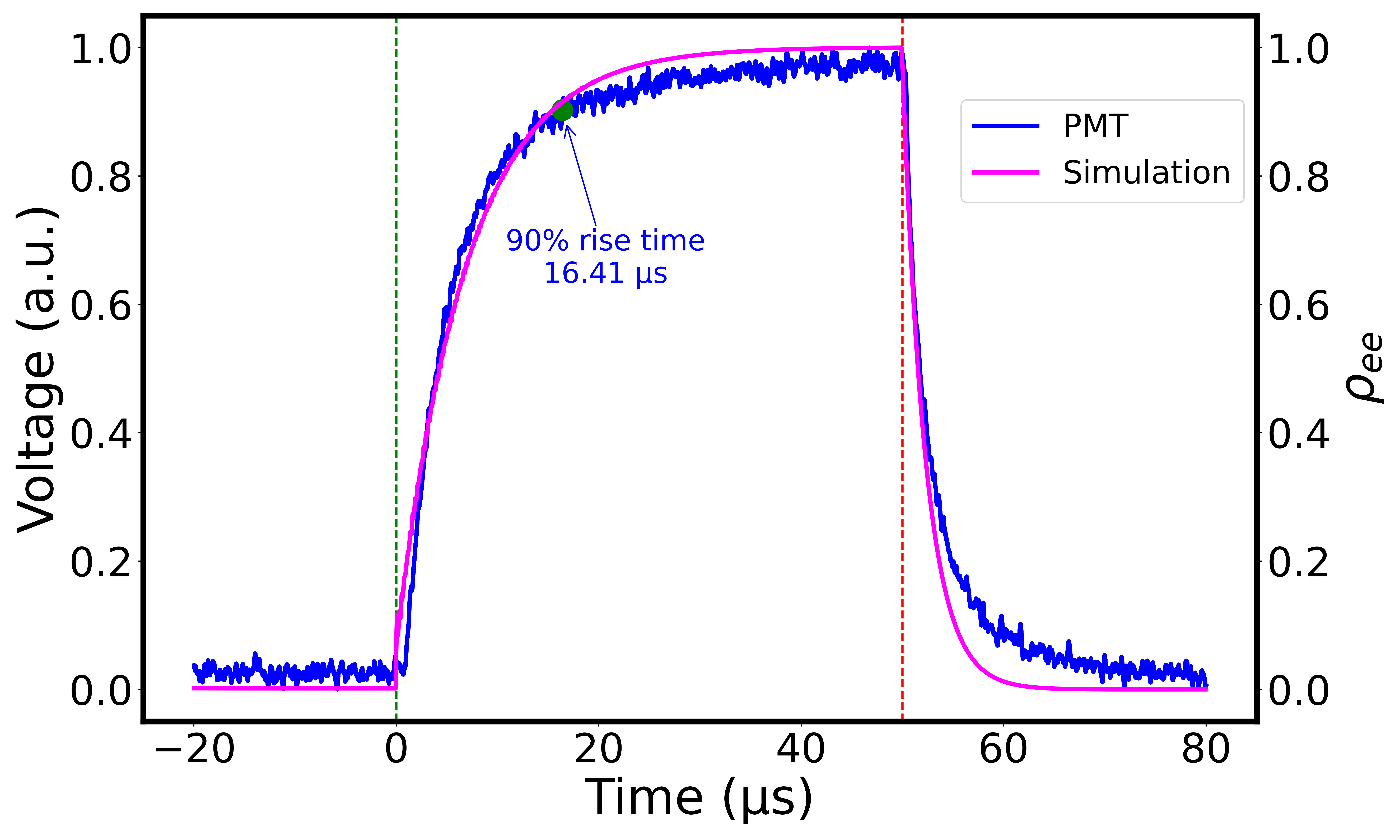}
\caption{\label{fig:epsart} (a) PMT voltage output in response to the mmWave pulse of 50 $\mu s$ duration. The on and off times of the pulse are indicated by vertical lines. The 16.41 $\mu s$ rise time is consistent with simulation of this system performed using similar parameters as in the experiment. Experimental data shown here is obtained using a 250 kHz low-pass filter at the PMT output and, averaged over 256 traces to reduce noise and get an accurate measurement of the rise time.}
\end{figure}

Next, we characterized the atom response-time-limited frame rate of our imaging system by measuring the fluorescence rise time in response to a sudden change in the RF field strength. We used a PIN diode switch with a fast-switching time ($\simeq$ 150 ns) to switch the input RF signal to the active multiplier, thereby creating a mmWave pulse. The resulting fluorescence signal was monitored using a PMT (PMT1001, Thorlabs) with a fast response time of 0.57 ns. As shown in Fig. 7, a mmWave pulse of 50 $\mu{s}$ duration is applied, and the corresponding fluorescence response is recorded. The fluorescence intensity was observed to reach 90$\%$ of its maximum value within 16.41 $\mu{s}$ after the RF pulse was triggered. Similar observations have been made in other studies while detecting RF pulses in probe beam transmission instead of fluorescence \cite{bohaichuk2022,sapiro2020}. We also conducted a numerical simulation using the Linblad master Eqn. (1) to study the time evolution of the atomic population in the intermediate excited state, $|e\rangle$ (i.e. $\rho_{ee}$) in response to the RF pulse \cite{miller2024}. Figure 7 shows a good agreement with our numerical simulation. The inverse of this rise time corresponds to the atom response-time-limited frame rate, which is approximately 60 kHz for our system, representing a significantly higher frame rate of operation over previously demonstrated fast mmWave imaging systems \cite{vakalis2021}. Such a high temporal resolution could potentially enable the capture of mmWave dynamics that could not be resolved with slower imaging techniques. Currently, in our mmWave imaging system, the frame rate is limited by the CCDs (i.e. 160 fps for  the Thorlabs CMOS camera, and 67 fps @ 2.2 MP resolution for the Tucsen CMOS camera). This, in principle, can be greatly improved using high-speed  camera.
\begin{figure}
\centering
\includegraphics[width=0.6\textwidth]{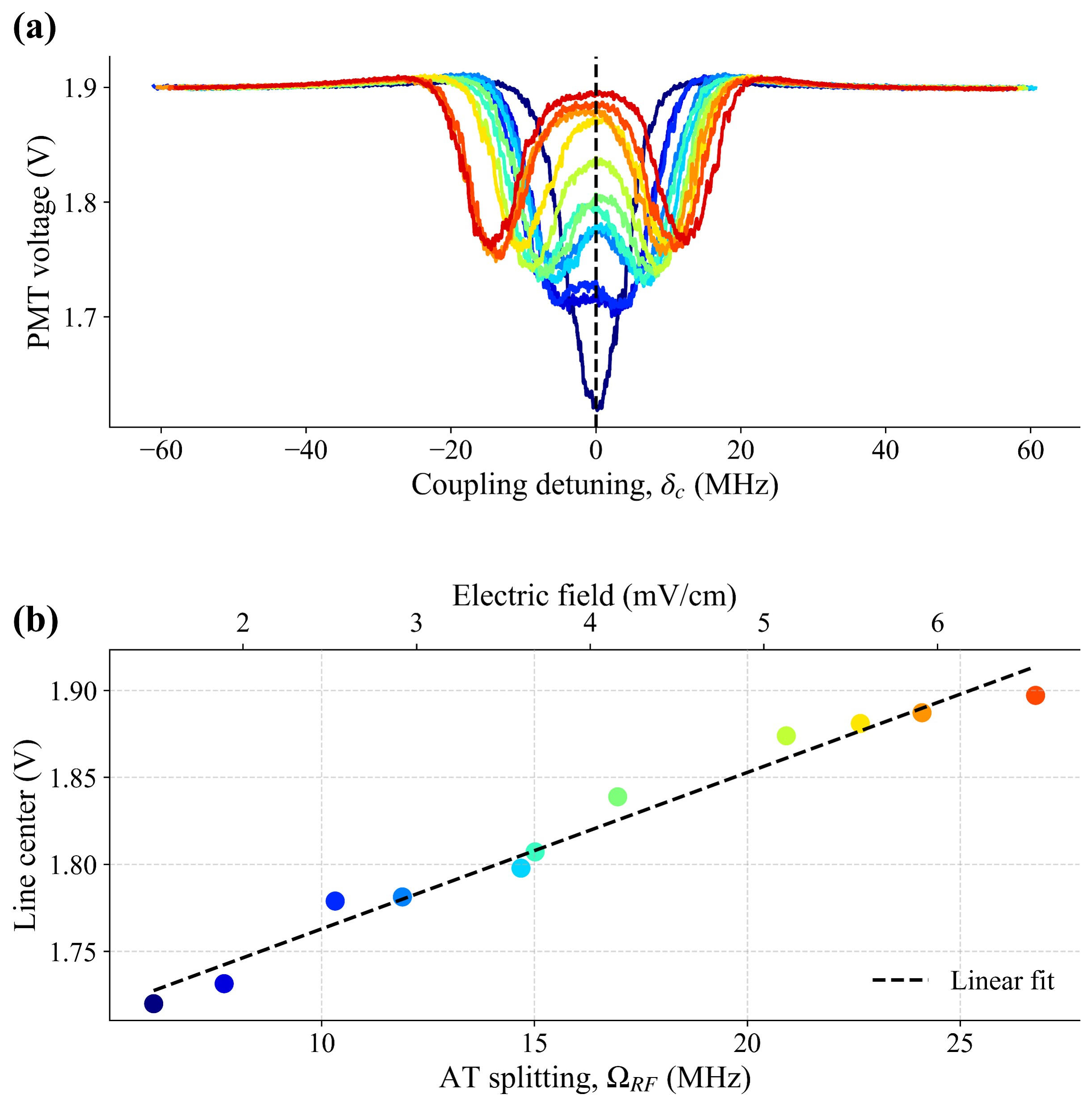}
\caption{\label{fig:epsart} (a) Resonances observed in fluorescence by scanning the coupling laser detuning, $\delta_c$ and applying the mmWave field. AT splitting of EIT dip is observed due to mmWave field. (b) PMT voltage measured at the line center (i.e. $\delta_c$ = 0) from figure (a) and plotted as a function of the AT splitting, $\Omega_{RF}$  (MHz) and the electric field (mV/cm) of the mmWave.}
\end{figure}

To characterize the sensitivity of our mmWave imaging system, we used a PMT first to measure the electric field amplitude of mmWave produced by the lens antenna with different settings of the variable attenuator (VA). The PMT was arranged in a ($2f-2f$) imaging configuration to collect fluorescence from the cell. In this configuration, the segmented area of the cell from which the fluorescence is collected has same size as that of PMT which is circular with a 8 mm diameter. Figure 8(a) shows resonances observed in fluorescence by scanning the coupling laser frequency (or detuning, $\delta_c$) and applying different amounts of mmWave (or RF) field using the VA. In the absence of the RF field, a single resonance dip due to EIT is observed. The AT splitting of the EIT dip is observed when the RF field is applied. Increasing the RF field strength increases the AT splitting, $\Omega_{RF}$, which is directly proportional to the electric field amplitude. Thus, we measured the AT splitting to measure the electric field strength [16]. The splitting causes the PMT voltage output (indicative of fluorescence) at the line center (i.e. $\delta_c$ = 0) to change. Figure 8(b) shows the measured voltage values at the line center [from Fig. 8(a)] plotted as a function of the AT splitting, $\Omega_{RF}$  (MHz) [and the electric field (mV/cm) of the mmWave which is calculated from AT splitting, $\Omega_{RF}$]. A linear fit to this data gives an estimated slope factor of approximately 111 MHz/V. This is a typical characteristic that we would expect to see from Region IV in Fig. 2 that we discussed in Sec. II. We used this slope factor to calculate the minimum detectable electric field strength and sensitivity of the PMT for this mmWave electric field measurement, by estimating the voltage noise in the signal. The voltage noise is estimated via two equivalent methods. We locked both the probe and coupling lasers to the line center and took a one second  measurement of fluorescence using the PMT with no applied RF field. We then calculated the RMS voltage noise, $\Delta V_{RMS}$ to be 2.7 mV. Alternatively, we also performed an FFT of the one second fluorescence measurement to calculate the total noise power in the signal over a 5 kHz bandwidth (assuming the noise to be dominant in the low frequency range) and calculated $\Delta V_{RMS}$ to be similar as before. Using the previously estimated slope factor of 111 MHz/V, this corresponds to a sensitivity $\delta{E}$ $\simeq$ 475 $\frac{\mu{V}}{cm\sqrt{Hz}}$ using the PMT for electric field measurement.

We then used a similar approach to estimate the sensitivity in mmWave imaging using the CCD camera. To determine the slope factor in this case, we recorded the pixel gray value in a single region of the image by applying a mmWave with a known field strength [already measured from Fig. 8(b)] using the variable attenuator. The image in this case was acquired using a 5 ms exposure time and averaged over 200 frames (with a total integration time of one second) to reduce noise in recording the gray value. 
\begin{figure}
\centering
\includegraphics[width=0.9\textwidth]{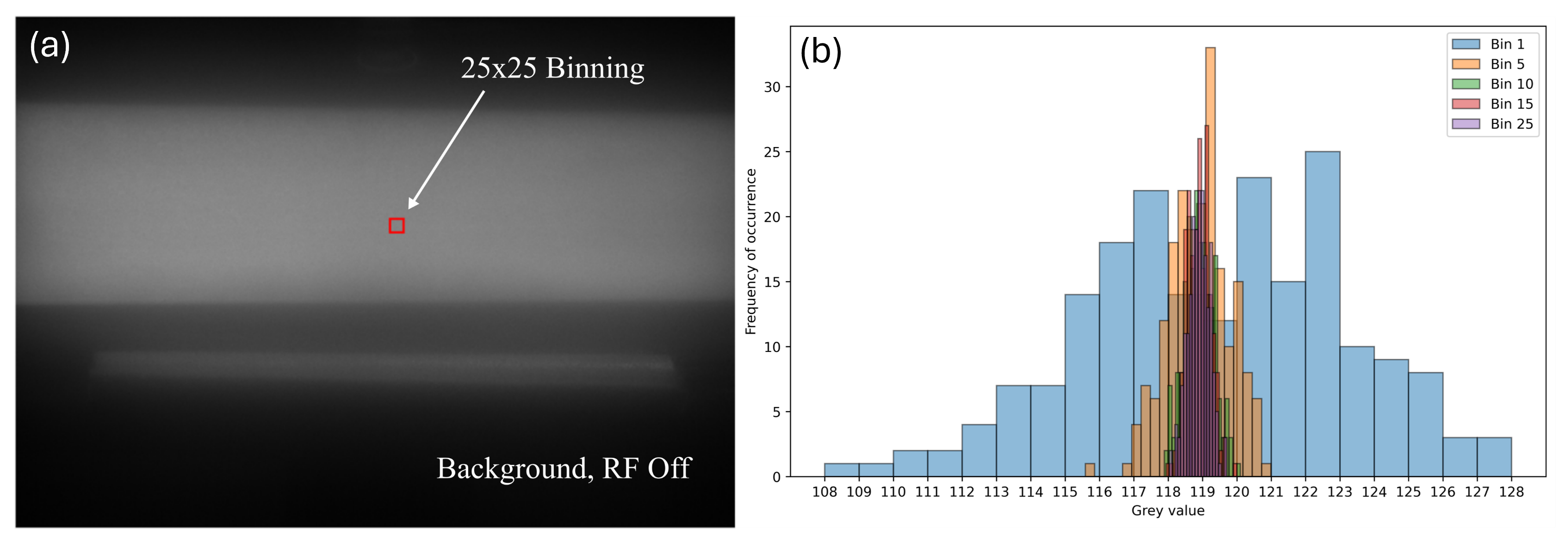}
\caption{\label{fig:epsart} (a) Image with background fluorescence showing a square region within the image chosen for CCD noise measurement. (b) Histograms of pixel gray values corresponding to different binning sizes from Bin 1 (single pixel) to Bin 25 (25$\times$25 pixels).}
\end{figure}
The slope factor corresponding to the ratio of field strength to gray value is then determined similar to that for the PMT as a function of AT splitting. The CCD noise is estimated without applying the mmWave and by finding the standard deviation of the pixel gray value (in a single region of the image) across 200 acquired frames. Fig. 9(a) shows the image with the background fluorescence and a square region chosen within this image (‘red square’) for CCD noise measurement. We also used the binning option to define this region from a single pixel to bins of 25 x 25 pixels. Fig. 9(b) shows histograms of gray values with different binning options. By increasing the bin size, the width of gray value distribution reduces, indicating a reduction in CCD noise. Since the spatial resolution in imaging in our system is limited by the mmWave  wavelength (i.e. $\lambda_{RF}$ = 3.28 mm), it is advantageous to use binning in image acquisition as it reduces CCD noise without sacrificing the spatial resolution. In our case, a binning of 25 x 25 pixels corresponds to a region of size 300 $\mu{m}$ x 300 $\mu{m}$ in the cell, which is well below the mmWave wavelength, and therefore, can be safely chosen for reducing the camera noise. We then measured the camera noise by estimating the standard deviations from the histograms shown in Fig. 9(b) and used the slope factor to calculate the single pixel minimum detectable field to be $E_{min}$ $\simeq$ 7.3 mV/cm. This is found to improve significantly to $E_{min}$ $\simeq$ 600 $\frac{\mu V}{cm}$ with 25 x 25 binning in imaging. Imaging sensitivity can be further improved by using a camera with low readout noise ($\leq{1}$ $e^-$ RMS) and a larger binning option. Hardware binning on the camera will also make the readout faster. We will explore this further in future.

To estimate the fundamental sensitivity limits of our system, we consider the theoretical noise floor due to photon shot noise (PSN) and atom projection noise (apn) \cite{fancher2021,tu2024,fan2015}. Using the measured PMT voltage in the $2f-2f$ imaging configuration at the line center, we estimated the mean photon flux to be $\Phi \simeq 6 \times 10^{12}$ photons/s which is obtained by converting the measured PMT voltage to photon flux based on the PMT parameters specified by the manufacturer. Since the photon shot noise can be modeled as $\sqrt{\Phi}$, we used a shot noise estimate of approximately $2.45 \times 10^6$, which is then converted back into a voltage shot noise $\Delta V\simeq$ 801 nV. Using the previously estimated slope factor $\frac{\partial\Omega_{RF}}{\partial V}$ = 111 $MHz/V$, we found the PSN-limited minimum detectable field $E_{min}^{psn}=[\Delta V \frac{1}{\frac{\partial V}{\partial\Omega_{RF}}}]\frac{\hbar}{d} = 137 \frac{nV}{cm}$ for PMT, where $d$ is the dipole moment of the $|r_1\rangle$ $\rightarrow$ $|r_2\rangle$ transition and $\hbar$ corresponds to reduced Planck constant. With a one second integration time, this corresponds to a PSN-limited sensitivity, $\delta E^{psn}$ of approximately 137 $\frac{nV}{cm\sqrt{Hz}}$.  A similar calculation is used to estimate the CCD photon shot noise. Using the parameters of CCD (Thorlabs CMOS camera) with high NIR sensitivity, full well capacity of 10,000 $e^-$, quantum efficiency of $45\%$ @ 780 nm, and gain of 1.8, the photon flux per pixel is found to be $\Phi_{pix}$=$1.13 \times 10^6$. For a 5 ms exposure time used in imaging, the fluctuation in gray value of the pixel due to photon shot noise is $\Delta P_{gv}$  = 1.55. Using the conversion $E_{min}^{psn} = [\Delta P_{gv} \frac{1}{\frac{\partial P_{gv}}{\partial\Omega_{RF}}}]\frac{\hbar}{d}$ with $\frac{\partial\Omega_{RF}}{\partial p_{gv}}$=1.176 MHz/$V_{gv}$, the PSN-limited minimum detectable electric field in a 5 ms exposure is found to be 2.8 mV/cm per pixel. This corresponds to a sensitivity, $\delta E^{psn}$ of 200 $\frac{\mu{V}}{cm\sqrt{Hz}}$ per pixel.

Similarly, the atom projection-noise-limited sensitivity is estimated using the relation $\delta E^{apn} = \frac{\hbar}{d\sqrt{T_2 N_{at}}}$, where $T_2$ is the coherence time, and $N_{at}$ is the number of atoms being measured. We measured the projection noise for PMT using the $2f-2f$ imaging configuration. In this case, $N_{at}$ is estimated to be $\approx$ $1.32 \times 10^6$ atoms using an atomic density of $1.32 \times 10^{16}$  atoms$/m^3$ for the room-temperature cell, and a collection volume of  $\approx$ $10^{-8} m^3$. The dominant limiting factor on  $T_2$ in our system is the laser linewidth which is conservatively taken to be 1 MHz, leading to a projection-noise-limited minimum detectable electric field of $E_{min}^{apn} \simeq 133 pV/cm$ and corresponding sensitivity of 133 $\frac{pV}{cm\sqrt{Hz}}$ for one second integration time. This is approximately three orders of magnitude higher than photon shot-noise-limited sensitivity for the PMT. A similar estimate of projection noise limited sensitivity is done for CCD (Thorlabs CMOS camera) using the volume of a single pixel in the image plane to be $3.02 \times 10^{-1}   m^3$ so that $N_{at}$  $\approx$ 4015 atoms. This leads to a minimum detectable electric field of 24 nV/cm and sensitivity of 1.71 $\frac{nV}{cm\sqrt{Hz}}$ for a 5 ms exposure time. The sensitivity 7.3 $\frac{mV}{cm\sqrt{Hz}}$ of our mmWave imaging system is closer to the photon shot noise limit. Deviation from this fundamental photon shot noise limit in the experiment is primarily due to Doppler broadening of the EIT linewidth which is $\approx$ 10 MHz.

Many relatively minor improvements could be made to our system to further improve performance. For example, we could extend this to a larger imaging area using a larger diameter cell, and by further expanding the probe and coupling beams. Although this may come at the expense of needing more laser power in the coupling beam, condensing the beams in the z-dimension and reflecting the coupling beam off dichroic mirrors may make this a viable option. A repump beam at 795 nm could be combined with the coupling beam to further improve sensitivity of our imaging approach \cite{prajapati2021}. We have also demonstrated that while the response time of 16.4 $\mu s$ is well beyond the current standard for mmWave imaging \cite{vakalis2021}, it can be further improved upon using optimized probe and coupling Rabi frequencies. Limitations on the frame rate in this work are due to the limited sensitivity and low frame rate of operation of the CCD used in imaging. Commercially available high-speed camera can be used to potentially allow high frame rate of operation in mmWave imaging. A high-speed lock-in camera (heliCam, Heliotis) can also be used to realize background-free imaging with high frame rates. Our future plans are to implement these improvements and attempt imaging of mmWave reflecting from large size objects at stand-off distance.
\section{\label{sec:level2}CONCLUSIONS}
We have demonstrated a mmWave imaging technique utilizing Autler-Townes splitting-induced fluorescence in a four-level Rydberg atomic system. Imaging the fluorescence from the intermediate atomic state at 780 nm effectively images the incident mmWave field. This gives an advantage to easily tune our system to other mmWave bands for imaging without changing the optical imaging. We have shown diffraction-limited imaging at 91.4 GHz with a rapid temporal response of 16.41 $\mu s$, showcasing the potential of our system for high-speed and high-resolution imaging. The ability to image vortex modes using phase plates highlights the versatility of our approach to image orbital angular momentum beams or other engineered beam profiles. We have demonstrated a sensitivity of $\approx$ 0.6  $\frac{mV}{cm\sqrt{Hz}}$ with 25 x 25 binning, which is close to the photon shot noise limit and suitable for active, real-time, stand-off mmWave imaging. We have also estimated the fundamental sensitivity limits of our system, by considering the photon shot noise and the atom projection noise. Our future work will focus on optimizing the imaging system for imaging larger objects in reflection with improved sensitivity closer to the fundamental sensitivity limits.

\begin{backmatter}
\bmsection{Acknowledgment}
This work is supported by the Center of Excellence on Advanced Quantum Sensing under the DoD grant W911NF2020276. We thank Kevin Cox and David Meyer at DEVCOM ARL for helpful discussions.    
\end{backmatter}

\bibliography{Rydberg-Optica}

\providecommand{\noopsort}[1]{}\providecommand{\singleletter}[1]{#1}%
\begin{thebibliography}{10}
\newcommand{\enquote}[1]{``#1''}

\bibitem{appleby2004}
R.~Appleby, \enquote{Passive millimetre--wave imaging and how it differs from terahertz imaging,} {\protect\JournalTitle{Phil. Trans. R. Soc. Lond.}} \textbf{362}, 379--393 (2004).

\bibitem{appleby2017}
R.~Appleby, D.~A. Robertson, and D.~Wikner, \enquote{Millimeter wave imaging: a historical review,} {\protect\JournalTitle{Proc. SPIE}} \textbf{10189}, 1018902 (2017).

\bibitem{wang2019}
Z.~Wang, T.~Chang, and H.-L. Cui, \enquote{Review of active millimeter wave imaging techniques for personnel security screening,} {\protect\JournalTitle{IEEE Access}} \textbf{7}, 148336--148350 (2019).

\bibitem{appleby2007}
R.~Appleby and R.~N. Anderton, \enquote{Millimeter-wave and submillimeter-wave imaging for security and surveillance,} {\protect\JournalTitle{Proc. IEEE}} \textbf{95}, 1683--1690 (2007).

\bibitem{nanzer2012}
J.~A. Nanzer, \emph{Microwave and millimeter-wave remote sensing for security applications} (Artech House, 2012).

\bibitem{vakalis2023}
S.~Vakalis, J.~R. Colon-Berrios, D.~Chen, and J.~A. Nanzer, \enquote{Non-destructive imaging of defects using non-cooperative 5g millimeter-wave signals,} {\protect\JournalTitle{Sensors}} \textbf{23}, 6421 (2023).

\bibitem{murakami2024}
H.~Murakami, T.~Fukuda, H.~Otera, \emph{et~al.}, \enquote{Development of a high-sensitivity millimeter-wave radar imaging system for non-destructive testing,} {\protect\JournalTitle{Sensors}} \textbf{24}, 4781 (2024).

\bibitem{mukherjee2019}
S.~Mukherjee, L.~Udpa, S.~Udpa, \emph{et~al.}, \enquote{A time reversal-based microwave imaging system for detection of breast tumors,} {\protect\JournalTitle{IEEE Trans. Microw. Theory Tech.}} \textbf{67}, 2062--2075 (2019).

\bibitem{mirbeik2019}
A.~Mirbeik-Sabzevari and N.~Tavassolian, \enquote{Tumor detection using millimeter-wave technology: Differentiating between benign lesions and cancer tissues,} {\protect\JournalTitle{IEEE Microw. Mag.}} \textbf{20}, 30--43 (2019).

\bibitem{patel2016}
V.~M. Patel, J.~N. Mait, D.~W. Prather, and A.~S. Hedden, \enquote{Computational millimeter wave imaging: problems, progress, and prospects,} {\protect\JournalTitle{IEEE Signal Processing Magazine}} \textbf{33}, 109--118 (2016).

\bibitem{fancher2021}
C.~T. Fancher, D.~R. Scherer, M.~C.~S. John, and B.~L.~S. Marlow, \enquote{Rydberg atom electric field sensors for communications and sensing,} {\protect\JournalTitle{IEEE Trans. Quant. Eng.}} \textbf{2}, 1--13 (2021).

\bibitem{zhang2024}
H.~Zhang, Y.~Ma, K.~Liao, \emph{et~al.}, \enquote{Rydberg atom electric field sensing for metrology, communication and hybrid quantum systems,} {\protect\JournalTitle{Science Bulletin}} \textbf{69}, 1515--1535 (2024).

\bibitem{holloway2014}
C.~L. Holloway, J.~A. Gordon, S.~Jefferts, \emph{et~al.}, \enquote{Broadband rydberg atom-based electric-field probe for si-traceable, self-calibrated measurements,} {\protect\JournalTitle{IEEE Trans. Ant. Prop.}} \textbf{62}, 6169--6182 (2014).

\bibitem{tu2024}
H.-T. Tu, K.-Y. Liao, H.-L. Wang, \emph{et~al.}, \enquote{Approaching the standard quantum limit of a rydberg-atom microwave electrometer,} {\protect\JournalTitle{Science Advances}} \textbf{10}, eads0683 (2024).

\bibitem{meyer2018}
D.~H. Meyer, K.~C. Cox, F.~K. Fatemi, and P.~D. Kunz, \enquote{Digital communication with rydberg atoms and amplitude-modulated microwave fields,} {\protect\JournalTitle{Appl. Phys. Lett.}} \textbf{112}, 211108 (2018).

\bibitem{deb2018}
A.~Deb and N.~Kj{\ae}rgaard, \enquote{Radio-over-fiber using an optical antenna based on rydberg states of atoms,} {\protect\JournalTitle{Appl. Phys. Lett.}} \textbf{112}, 211106 (2018).

\bibitem{gordon2014}
J.~A. Gordon, C.~L. Holloway, A.~Schwarzkopf, \emph{et~al.}, \enquote{Millimeter wave detection via autler-townes splitting in rubidium rydberg atoms,} {\protect\JournalTitle{Appl. Phys. Lett.}} \textbf{105}, 024104 (2014).

\bibitem{yuan2023}
J.~Yuan, T.~Jin, L.~Xiao, \emph{et~al.}, \enquote{A \uppercase{R}ydberg atom-based receiver with amplitude modulation technique for the fifth-generation millimeter-wave wireless communication,} {\protect\JournalTitle{IEEE Ant. Wire. Prop. Lett.}} \textbf{22}, 2580--2584 (2023).

\bibitem{legaie2024}
R.~Legaie, G.~Raithel, and D.~A. Anderson, \enquote{A millimeter-wave atomic receiver,} {\protect\JournalTitle{AVS Quant. Sci.}} \textbf{6} (2024).

\bibitem{borowka2024}
S.~Bor{\'o}wka, W.~Krokosz, M.~Mazelanik, \emph{et~al.}, \enquote{Rydberg-atom-based system for benchmarking millimeter-wave automotive radar chips,} {\protect\JournalTitle{Physical Review Applied}} \textbf{22}, 034067 (2024).

\bibitem{fan2014}
H.~Fan, S.~Kumar, R.~Daschner, \emph{et~al.}, \enquote{Subwavelength microwave electric-field imaging using rydberg atoms inside atomic vapor cells,} {\protect\JournalTitle{Opt. Lett.}} \textbf{39}, 3030--3033 (2014).

\bibitem{wade2017}
C.~G. Wade, N.~{\v{S}}ibali{\'c}, N.~R. De~Melo, \emph{et~al.}, \enquote{Real-time near-field terahertz imaging with atomic optical fluorescence,} {\protect\JournalTitle{Nature Photonics}} \textbf{11}, 40--43 (2017).

\bibitem{downes2020}
L.~A. Downes, A.~R. MacKellar, D.~J. Whiting, \emph{et~al.}, \enquote{Full-field terahertz imaging at kilohertz frame rates using atomic vapor,} {\protect\JournalTitle{Phys. Rev. X}} \textbf{10}, 011027 (2020).

\bibitem{downes2023}
L.~A. Downes, L.~Torralbo-Campo, and K.~J. Weatherill, \enquote{A practical guide to terahertz imaging using thermal atomic vapour,} {\protect\JournalTitle{New J. Phys.}} \textbf{25}, 035002 (2023).

\bibitem{schlossberger2024}
N.~Schlossberger, T.~McDonald, K.~Su, \emph{et~al.}, \enquote{Two-dimensional imaging of electromagnetic fields via light sheet fluorescence imaging with rydberg atoms,} {\protect\JournalTitle{arXiv preprint arXiv:2412.12568}}  (2024).

\bibitem{manzano2020}
D.~Manzano, \enquote{A short introduction to the lindblad master equation,} {\protect\JournalTitle{AIP Adv.}} \textbf{10}, 025106 (2020).

\bibitem{miller2024}
B.~N. Miller, D.~H. Meyer, T.~Virtanen, \emph{et~al.}, \enquote{Rydiqule: A graph-based paradigm for modeling rydberg and atomic sensors,} {\protect\JournalTitle{Comp. Phys. Commun.}} \textbf{294}, 108952 (2024).

\bibitem{sibalic2017}
N.~{\v{S}}ibali{\'c}, J.~D. Pritchard, C.~S. Adams, and K.~J. Weatherill, \enquote{Arc: An open-source library for calculating properties of alkali rydberg atoms,} {\protect\JournalTitle{Comp. Phys. Commun.}} \textbf{220}, 319--331 (2017).

\bibitem{anisimov2011}
P.~M. Anisimov, J.~P. Dowling, and B.~C. Sanders, \enquote{Objectively discerning autler-townes splitting from electromagnetically induced transparency,} {\protect\JournalTitle{Phys. Rev. Lett.}} \textbf{107}, 163604 (2011).

\bibitem{yao2022}
J.~Yao, Q.~An, Y.~Zhou, \emph{et~al.}, \enquote{Sensitivity enhancement of far-detuned rf field sensing based on rydberg atoms dressed by a near-resonant rf field,} {\protect\JournalTitle{Opt. Lett.}} \textbf{47}, 5256 (2022).

\bibitem{vargas2022}
P.~E. Vargas, E.~Meyer, F.~Chiappini, \emph{et~al.}, \enquote{Characterization of the dielectric properties of commercially available low-loss uv-curable resins from 60 ghz to 90 ghz,}  (IEEE, 2022), pp. 278--281.

\bibitem{sahin2019}
S.~Sahin, N.~K. Nahar, and K.~Sertel, \enquote{Dielectric properties of low-loss polymers for mmw and thz applications,} {\protect\JournalTitle{J. Infra. Milli.Tera. Waves}} \textbf{40}, 557--573 (2019).

\bibitem{schemmel2014}
P.~Schemmel, G.~Pisano, and B.~Maffei, \enquote{Modular spiral phase plate design for orbital angular momentum generation at millimetre wavelengths,} {\protect\JournalTitle{Opt. Exp.}} \textbf{22}, 14712--14726 (2014).

\bibitem{downes2022}
L.~A. Downes, D.~J. Whiting, C.~S. Adams, and K.~J. Weatherill, \enquote{Rapid readout of terahertz orbital angular momentum beams using atom-based imaging,} {\protect\JournalTitle{Opt. Lett.}} \textbf{47}, 6001--6004 (2022).

\bibitem{ge2017}
X.~Ge, R.~Zi, X.~Xiong, \emph{et~al.}, \enquote{Millimeter wave communications with oam-sm scheme for future mobile networks,} {\protect\JournalTitle{IEEE J. Sel. Area. Commun.}} \textbf{35}, 2163--2177 (2017).

\bibitem{yan2014}
Y.~Yan, G.~Xie, M.~P. Lavery, \emph{et~al.}, \enquote{High-capacity millimetre-wave communications with orbital angular momentum multiplexing,} {\protect\JournalTitle{Nature Commun.}} \textbf{5}, 4876 (2014).

\bibitem{BeamLab}
\enquote{Beamlab: Matlab toolboxes for optical simulations,} .

\bibitem{bohaichuk2022}
S.~M. Bohaichuk, D.~Booth, K.~Nickerson, \emph{et~al.}, \enquote{Origins of rydberg-atom electrometer transient response and its impact on radio-frequency pulse sensing,} {\protect\JournalTitle{Phys. Rev. Appl.}} \textbf{18}, 034030 (2022).

\bibitem{sapiro2020}
R.~Sapiro, G.~Raithel, and D.~Anderson, \enquote{Time dependence of rydberg eit in pulsed optical and rf fields,} {\protect\JournalTitle{J. Phys. B: Atomic, Molecular and Optical Physics}} \textbf{53}, 094003 (2020).

\bibitem{vakalis2021}
S.~Vakalis, D.~Chen, and J.~A. Nanzer, \enquote{Millimeter-wave imaging at 652 frames per second,} {\protect\JournalTitle{IEEE J. Microw.}} \textbf{1}, 738--746 (2021).

\bibitem{fan2015}
H.~Fan, S.~Kumar, J.~Sedlacek, \emph{et~al.}, \enquote{Atom based rf electric field sensing,} {\protect\JournalTitle{J. Phys. B: Atomic, Molecular and Optical Physics}} \textbf{48}, 202001 (2015).

\bibitem{prajapati2021}
N.~Prajapati, A.~K. Robinson, S.~Berweger, \emph{et~al.}, \enquote{Enhancement of electromagnetically induced transparency based rydberg-atom electrometry through population repumping,} {\protect\JournalTitle{Applied Physics Letters}} \textbf{119}, 214001 (2021).

\end{thebibliography}

\end{document}